\documentclass[11pt,a4paper,twoside,groupcitations]{article}
\usepackage[T1]{fontenc}
\usepackage[ansinew]{inputenc}
\usepackage[english]{babel}
\usepackage{amsfonts}
\usepackage{amsmath}
\usepackage{bm}
\usepackage{array}
\usepackage{amsthm}
\usepackage{amssymb}
\usepackage{graphicx}
\usepackage{subfigure}
\usepackage{braket}
\usepackage{eucal}
\usepackage{verbatim}
\usepackage[table]{xcolor}
\usepackage{caption}
\usepackage{cite}
\usepackage{textcomp}
\usepackage{multicol}
\usepackage{tikz}
\usetikzlibrary{positioning,arrows}
\usetikzlibrary{decorations.pathmorphing}
\usetikzlibrary{decorations.markings}
\usetikzlibrary{calc,decorations.markings}
\usetikzlibrary{arrows,shapes}
\usetikzlibrary{matrix,arrows}
\usepackage{pgfplots}
\usepackage{xparse}
\definecolor{jade}{HTML}{00A86B}
\newcommand{\be}{\begin{eqnarray}}
\newcommand{\ee}{\end{eqnarray}}

\renewcommand{\d}{\mbox{${\rm d}$}} 

\newcommand{\gn}{G_{\rm N}}
\newcommand{\rh}{r_{\rm H}}

\title{\bf Star equilibrium: from BNG to TOV}
\author{Roberto~Casadio$^{ab}$\thanks{E-mail: casadio@bo.infn.it}
$\ $
and
Octavian Micu$^c$\thanks{E-mail: octavian.micu@spacescience.ro}
\\
\\
$^a${\em Dipartimento di Fisica e Astronomia, Universit\`a di Bologna}
\\
{\em via Irnerio~46, 40126 Bologna, Italy}
\\
\\
$^b${\em I.N.F.N., Sezione di Bologna, I.S.~FLAG}
\\
{\em viale B.~Pichat~6/2, 40127 Bologna, Italy}
\\
\\
$^c${\em Institute of Space Science - Subsidiary of INFLPR}
\\
{\em P.O. Box MG-23, RO-077125 Bucharest-Magurele, Romania}
}
\begin{document}
\maketitle
\begin{abstract}
We study the role of the equilibrium equation in bootstrapped Newtonian gravity (BNG) 
by including terms inspired by the post-Newtonian expansion of the
Tolman-Oppenheimer-Volkov (TOV) equation.
We then compare (approximate) BNG solutions for homogenous stars with their Newtonian
and General Relativistic exact solutions.
Regardless of the additional terms from the conservation equation, BNG stars do not exhibit
a Buchdahl limit. 
However, specific extra terms added to this equation can cause the pressure
to become negative inside stars with compactness smaller than the critical
values for BNG black hole formation.   
\end{abstract}
\section{Introduction}
\setcounter{equation}{0}
\label{S:BN}
We live in a time when more and more experiments are aimed at testing (directly or indirectly)
the gravitational interaction in its strong-field regime.
Indeed there are many mysteries surrounding the nature of the extremely compact objects that
would source these strong fields.
A key question is whether space-time singularities exist behind black hole event
horizons, as predicted by general relativity (GR)~\cite{Hawking:1973uf}, or whether an alternative
description is needed in this regime that can lead to regular interior solutions
(see, e.g.~Ref.~\cite{Bambi:2023try}), for example to effectively account for quantum effects
in the classical theory.
With this in mind, bootstrapped Newtonian gravity (BNG) was introduced in Ref.~\cite{Casadio:2018qeh} and 
subsequently developed in a series of works~\cite{Casadio:2019cux,Casadio:2020kbc,Casadio:2020ueb,
Casadio:2019pli,Casadio:2021gdf,Casadio:2022gbv,Casadio:2022pme} as a simple theoretical
framework to investigate the role that non-linearities added to the Newtonian theory could play
in the interior of very compact objects and even possibly black holes.
One of its main motivations was exactly to find if there could exist non-singular configurations for the
matter trapped behind the event horizon of a black hole in a description of gravity that is not quite the
full-fledged GR.
BNG is viewed as a bottom-up approach, that starts from the Newtonian theory and incorporates
several additional (non-linear) terms similar to those that appear at leading orders in the weak-field
expansion of GR.
However, all additional terms are treated on equal footing from the outset, and we regard BNG as a
distinct alternative description of compact objects.
\par
The additional non-linear terms add a considerable amount of complexity.
As a result, it becomes usually impossible to find analytic solutions. 
In BNG, the model for a compact object is governed by a system of two differential equations,
namely the Euler-Lagrange equation for the gravitational potential and the conservation equation,
which are solved by imposing the corresponding boundary conditions.
While the Euler-Lagrange equation obtained from the action that includes the non-linear terms
has remained unchanged throughout the development of the BNG, higher levels of complexity were
achieved in successive increments by taking into account extra terms in the equation that
is responsible for the equilibrium between the gravitational pull and the pressure. 
Initially the simplest Newtonian conservation equation was employed.
However, the pressure was shown to become very large in the compactness regime of interest, and the
pressure was then added to the density, in what might have seemed an ad-hoc extension.
In order to achieve a more comprehensive understanding of what our model of compact objects
can lead to, we will here add all of the first order terms resulting from the expansion of the TOV
equation in the Newton constant and pressure.
Besides extending the model, this also works as a check which shows that the pressure term added
previously is found among these extra first order terms.
\par
The BNG is built upon a Lagrangian that incorporates several additional terms beyond
those found in the Lagrangian for the Newtonian potential,
\be
L_{\rm N}[V]
=
-4\,\pi
\int_0^\infty
r^2 \,\d r
\left[
\frac{\left(V'\right)^2}{8\,\pi\,\gn}
+\rho\,V
\right]
\ ,
\label{LagrNewt}
\ee
which leads to the well known Poisson equation in spherical coordinates, with primes denoting
derivatives with respect to $r$.
More exactly, the BNG Lagrangian from Ref.~\cite{Casadio:2019cux} is given by 
\be 
L[V]
&\!\!=\!\!&
L_{\rm N}[V]
-4\,\pi
\int_0^\infty
r^2\,\d r
\left[
q_V\,\mathcal{J}_V\,V
+
q_p\,\mathcal{J}_p\,V
+
q_\rho\, \mathcal{J}_\rho \left(\rho+q_p\,\mathcal{J}_p\right)
\right]
\nonumber
\\
&\!\!=\!\!&
-4\,\pi
\int_0^\infty
r^2\,\d r
\left[
\frac{\left(V'\right)^2}{8\,\pi\,\gn}
\left(1-4\,q_V\, V\right)
+\left(\rho+\,q_p\,p\right)
V
\left(1-2\,q_\rho\, V\right)
\right]
\ ,
\label{LagrV}
\ee
with $V=V(r)$ representing the potential for static spherically symmetric objects,
and primes denote derivatives with respect to the radial coordinate $r$.
Without the additional terms proportional to $q_V$, $q_p$ and $q_\rho$,
the potential $V$ is the same as the Newtonian potential $V_{\rm N}$ satisfying
the field equations derived from Eq.~\eqref{LagrNewt} for an object of matter
density $\rho=\rho(r)$.
Note that when comparing with GR expressions, the radius $r$ in the above
Lagrangian should be viewed as the ``harmonic'' radial coordinate (which leads
to the Poisson equation for the Newtonian potential from the weak-field expansion
of the Einstein equations~\cite{weinberg}).
\par 
The term proportional to $q_V$ in the integral~\eqref{LagrV} should be
understood as the gravitational self-coupling sourced by the gravitational energy
$U_{\rm N}$ per unit volume
\be
\mathcal{J}_V
\simeq
\frac{\d U_{\rm N}}{\d \mathcal{V}} 
=
-\frac{\left[ V'(r) \right]^2}{2\,\pi\,\gn}
\ .
\label{JV}
\ee
The BNG was designed to be applicable to highly compact objects, with the compactness defined as
\be 
X\equiv \frac{\gn\, M}{R}
\ ,
\ee
where $M$ represents the Arnowitt-Deser-Misner~(ADM)-like mass~\cite{Arnowitt:1959ah},
which should be understood as the mass calculated by studying orbits around the
star~\cite{Casadio:2021gdf,DAddio:2021xsu}, and $R$ is the star radius.
For $X\geq 0.1$, the static pressure $p=p(r)$ was found to be no longer negligible~\cite{Casadio:2018qeh}. 
A corresponding potential energy term $U_p$ was then introduced in Ref.~\cite{Casadio:2019cux},
such that 
\be
\mathcal{J}_p
\simeq
-\frac{\d U_p}{\d \mathcal{V}} 
=
p
\ .
\label{JP}
\ee
The effect of the pressure term is to replace $\rho \to \rho+q_p\,p$ in the Lagrangian,
with the dimensionless coupling $q_p$ introduced to formally interpolate between the relativistic
limit $q_p\to 1$ and the non-relativistic limit $q_p\to 0$.
The inclusion of the pressure by effectively shifting the density follows along the lines of the
definition of the Tolman mass~\cite{tolman}.
\par
The last term in the Lagrangian in Eq.~\eqref{LagrV} is a higher-order term, namely
\be
\mathcal{J}_\rho
=
-2\,V^2
\ ,
\ee
which couples with the matter source.
In fact, all of the coupling constants $q_V$, $q_p$ and $q_\rho$ were specifically introduced to study
the different regimes of the BNG when they vary between zero and one.
The Newtonian regime is recovered in the limit $q_V=q_p=q_\rho\to 0$, as the only remaining term in
Eq.~\eqref{LagrV} is the Lagrangian for the Newtonian potential.
Each of the additional terms in Eq.~\eqref{LagrV} was discussed in more detail in Ref.~\cite{Casadio:2019cux}
and the effects of the coupling $q_\rho$ were further investigated in Ref.~\cite{Casadio:2019pli}.
\par
Varying the action~\eqref{LagrV} with respect to $V$ yields the Euler-Lagrange equation
\be
\triangle V
=
4\,\pi\,\gn\left(\rho+q_p\,p\right)
\frac{1-4\,q_\rho\,V}{1-4\,q_V\,V}
+
\frac{2\,q_V\left(V'\right)^2}
{1-4\,q_V\,V}
\ ,
\label{EOMV}
\ee
where $\triangle$ is the Laplacian operator in spherical coordinates.
\par 
The conservation equation for the energy-momentum tensor of a spherically symmetric fluid
in GR is given by the Tolman-Oppenheimer-Volkov (TOV) equation~\cite{Tolman:1939jz,Oppenheimer:1939ne},
which follows from the Einstein equations without extra assumptions [see Appendix~\ref{A:NpN}].
Instead, one must impose an equilibrium equation between the pressure and the gravitational pull
in Newtonian gravity.
The situation is similar in the BNG, where the equilibrium equation must also be imposed beside the
field equation~\eqref{EOMV} that determines the potential.
In the original proposal~\cite{Casadio:2018qeh}, we assumed the Newtonian conservation equation,
\be
p'
\simeq
-
\rho\,V'
\ .
\label{eqP_OLD}
\ee
Noting that some configuration involved very large values of $p$, we later added its contribution
to the density term~\cite{Casadio:2019cux},
so that
\be
p'
\simeq
-
\left( \rho + p\right) V'
\ .
\label{eqP_old}
\ee 
Using the above conservation equation, it was found in Ref.~\cite{Casadio:2019cux} that
there is no Buchdahl limit~\cite{Buchdahl:1959zz} for isotropic sources. 
\par
In this work, we perform a more systematic investigation of the effects of the conservation
equation and how the BNG can ``interpolate'' between the Newtonian regime and the full GR
picture in the equilibrium of compact  astrophysical objects.
For this purpose, we expand the TOV equation in the Newton constant and pressure
up to second order in Appendix~\ref{A:NpN}, and obtain the modified equilibrium equation
\be
p'
\simeq
-
\left( \rho + p\right) V' - 4\,\pi\,\gn\,r\, p\,\rho -2\,\rho\,V\,V'
\ ,
\label{eqp}
\ee 
where we set the coupling constants $q_V=q_p=q_\rho=\epsilon=1$.
In the following Sections, we will compare solutions of the field equation
\be
\triangle V
=
4\,\pi\,\gn\left(\rho+p\right)
+
\frac{2\left(V'\right)^2}
{1-4\,V}
\label{EOMV1}
\ee
supplemented by the conservation Eq.~\eqref{eqP_old} with solutions obtained by adding the 
additional terms in Eq.~\eqref{eqp}.
In particular, we will employ a homogeneous density profile, since this case can be analytically
solved both in Newtonian gravity and in GR [see Section~\ref{S:homo}], and analyse the effect
of adding the second term in the right hand side of Eq.~\eqref{eqp} in Section~\ref{S:addP}, respectively the 
complete Eq.~\eqref{eqp} in Section~\ref{S:addPV}.
Approximate analytic and numerical solutions for the potential and interior pressure are obtained in each case. 
These two solutions are compared by calculating the relative difference between them. The solutions for the pressure 
are also compared to the approximate GR pressure derived in Appendix~\ref{A:homo}. 
The main difference with respect to the previously analysed cases resulting from the addition of 
the last term from Eq.~\eqref{eqp} is that, while the approximate potential remains well behaved, 
the pressure inside these BNG stars becomes negative before they reach black hole compactness values. 
\section{Homogeneous spherical stars and BNG vacuum}
\label{S:homo}
\setcounter{equation}{0}
Isotropic stars of radius $r=R$ with homogeneous density
\be
\rho
=
\left\{
\begin{array}{ll}
\strut\displaystyle\frac{3\,M_0}{4\,\pi\,R^3}
\equiv 
\rho_0
\qquad
&
{\rm for}\ 0\le r< R
\\
\\
0
&
{\rm for}\ R<r
\end{array}
\right.
\label{HomDens}
\ee
are the simplest cases that can be studied analytically both in GR and Newtonian physics. 
The exact GR solutions are recalled in Appendix~\ref{A:homo} along with their
Newtonian approximation.
In particular, GR and Newtonian solutions in the outer vacuum are unique and do not
depend on the interior of the star but only on its mass:
the GR solution contains the ADM mass $\bar M=m(\bar R)$ obtained from Eq.~\eqref{massclass},
whereas the Newtonian solution contains the proper mass $M_0$ in Eq.~\eqref{HomDens}
given by the flat volume integral of the energy density.
\par
The vacuum solution in BNG is also unique and does not depend on the equilibrium
equation.
In fact, Eqs.~\eqref{eqP_OLD}-\eqref{eqp} are all trivially satisfied for $r>R$ where $\rho=p=0$,
while Eq.~\eqref{EOMV1} becomes
\be
\triangle V
=
\frac{2 \left(V'\right)^2}{1-4\,V}
\ . 
\label{EOMV0}
\ee
After fixing the integration constants so as to recover the Newtonian behaviour at infinity,
one finds the unique solution~\cite{Casadio:2018qeh}
\be
V_{\rm out}
=
\frac{1}{4}
\left[
1-\left(1+\frac{6\,\gn\,M}{r}\right)^{2/3}
\right]
\ .
\label{sol0}
\ee
The above expression at large $r$ yields the Newtonian term and the next-to-leading
order post-Newtonian term of order $\gn^2$, without any additional assumptions.
An important difference with respect to Newtonian gravity is that the 
ADM-like mass $M$ above does not equal the proper mass $M_0$ in Eq.~\eqref{HomDens},
which remains defined by
\be
M_0
=
4\,\pi
\int_0^{R}
r^2\,\d r\,\rho(r)
\ .
\ee
Note that $r$ above represents the radial distance from the centre in both Newtonian gravity and BNG,
so that $M_0$ is indeed the proper mass of the star in a flat space.~\footnote{The proper mass in GR is given by the
spatial integral of the density with the appropriate volume measure.}
\par
The potential must be smooth across the star surface located at $r=R$, which means
\be
V_{\rm in}(R)
=
V_{\rm out}(R)
\equiv
V_R
=
\frac{1}{4}
\left[
1-\left(1+6\,X\right)^{2/3}
\right]
\ ,
\label{VR} 
\ee
and
\be
V'_{\rm in}(R)
=
V'_{\rm out}(R)
\equiv
V'_R
=
\frac{X}
{R\left(1+6\, X\right)^{1/3}}
\ ,
\label{dVR}
\ee
where $V_{\rm in}=V(0<r<R)$ denotes the potential inside the source.
Finally, the potential must be regular at $r=0$,
\be
V_{\rm in}'(0)=0 
\ ,
\label{b0}
\ee
in order to accommodate smooth density profiles in the centre.
\section{Additional pressure term}
\setcounter{equation}{0}
\label{S:addP}
After including the extra term proportional to the pressure, the conservation equation becomes
\be
p'
\simeq
-
\left( \rho + p\right) V' - 4\,\pi\,\gn\,r\, p\,\rho
\ .
\label{eqp_1}
\ee 
Together with the field equation~\eqref{EOMV1} this forms a system of differential equations
which, along with the boundary conditions at $r=0$ and $r=R$, determine the gravitational
potential $V$ and the pressure $p$ given the homogenous density~\eqref{HomDens}.
Additionally, the dependence of the density on the ADM-like mass $M$ which results from
solving the system, determines $M$ in terms of the proper mass $M_0$ in Eq.~\eqref{HomDens}. 
Unfortunately, finding analytic solutions for a general value of the compactness $X$
proves to be an impossible task.
\par
Following previous works~\cite{Casadio:2019cux}, we start from an approximate analytic
solution $V_{\rm s}$ for the potential $V_{\rm in}$ obtained from a series expansion around $r=0$,
to wit
\be
V_{\rm s}
=
V_0
+
V_2\,r^2
\ ,
\label{Vs0}
\ee
where $V_0\equiv V_{\rm in}(0)<0$.
The term proportional to $r$ must vanish in order for the regularity condition~\eqref{b0}
to be fulfilled.
The same constraint eventually leads to the vanishing of all odd orders in $r$ in the
Taylor expansion about $r=0$.
\par
\begin{figure}[tp]
\centering
\includegraphics[width=8cm]{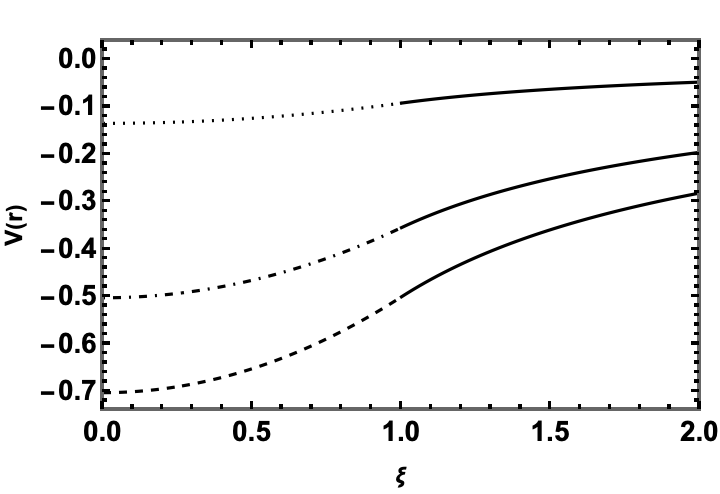}
\caption{Approximate potential~\eqref{Vins} for sources of compactness $X=0.1$ (dotted line),
$X=0.46$ (dot-dashed line) and $X=0.7$ (dashed lines) joined to the corresponding outer
potential~\eqref{sol0} for the same compactness (solid lines).}
\label{Fig1}
\end{figure}

The boundary conditions at the edge of the star from Eqs.~\eqref{VR} and \eqref{dVR}
allow us to fix the factors $V_0$, respectively $V_2$, and write the approximate potential as
\be
V_{\rm s}
=
-\frac{1+8\,X-\left(1+6\,X\right)^{1/3} }{4 \left(1+6\,X\right)^{1/3}}
+\frac{X\,\xi^2}{2\,\left(1+6\,X\right)^{1/3}}
\ ,
\label{Vins}
\ee
where $\xi\equiv r/R$.
It is important to remark that the potential inside the source is fully determined by the
matching conditions with the unique outer potential in this approximation.
Nonetheless, it will be shown later that this approximation fits well the numerical solutions
of Eqs.~\eqref{EOMV1} and~\eqref{eqp_1} when the compactness is not very large. 
The potential inside the source and its continuation in the outer vacuum are plotted
in Fig.~\ref{Fig1} for several values of the compactness $X$.
\par
For consistency, a similar approximation is employed for the pressure, namely $p\simeq p_{\rm s}$ with
\be
p_{\rm s}
=
p_0
+
p_2\,r^2
\ ,
\label{pin}
\ee
where, again, the term linear in $r$ must vanish to ensure the regularity condition~\eqref{b0}. 
Using the above approximate expressions for the potential and pressure in Eqs.~\eqref{EOMV1}
and~\eqref{eqp_1} yields
\be
p_{\rm s}
&\!\!=\!\!&
\frac{X^2\left(1 - \xi^2\right)}{8\,\pi\,\gn^3\,M^2}
\left[
\frac{3\,X}{(1+6\,X)^{1/3}}
- 
2\left(1-
\sqrt{1+\frac{3\,X\left[7\,X-4\,(1+6\,X)^{1/3}\right]}{4\,(1+6\,X)^{2/3}}}
\right)
\right]
\nonumber
\\
&\!\!\simeq\!\!&
\frac{3\,X^4\left(1 - \xi^2\right)}{8\,\pi\,\gn^3\,M^2}
\ ,
\label{pinsol}
\ee
where the last expression is the leading order term for $X\ll 1$.
\par
\begin{figure}[]
\centering
\includegraphics[width=8.2cm]{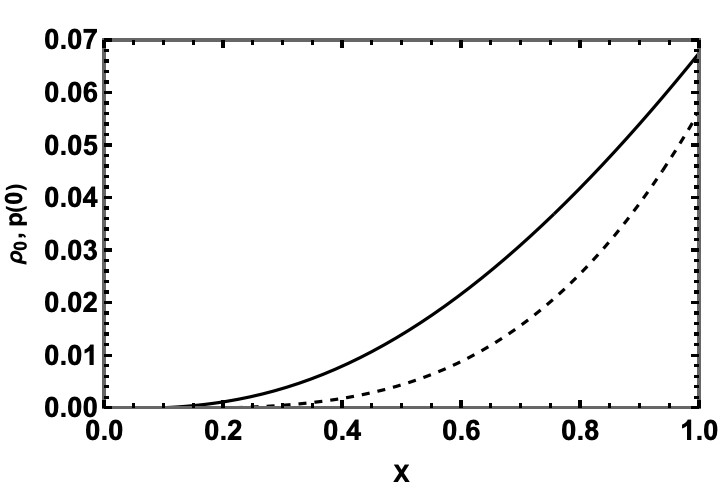}
\includegraphics[width=8cm]{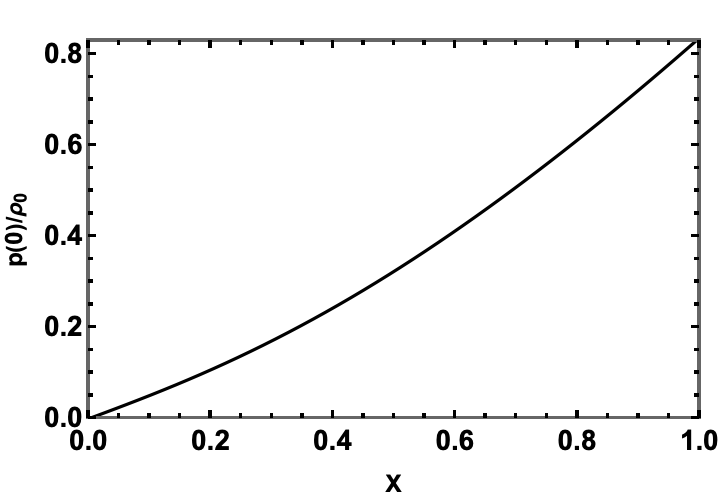}
\caption{Left panel: density (solid line) and approximate central pressure (dashed line) 
in units of $\gn^{-3}\,M^{-2}$.
Right panel: ratio of the central pressure and density.}
\label{p0andrho}
\end{figure}
\begin{figure}[]
\centering
\includegraphics[width=8cm]{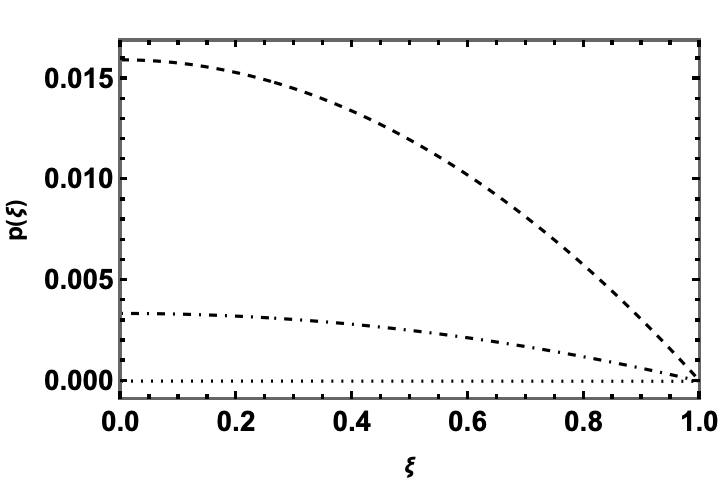}
\caption{Analytic approximation~\eqref{pinsol} for the pressure (in units of $\gn^{-3}\,M^{-2}$)
for compactness $X=0.1$ (dotted line), $X=0.46$ (dot-dashed line) and $X=0.7$ (dashed line).}
\label{pxi}
\end{figure}

Within this approximation, the density can be expressed in terms of the ADM-like mass as
\be
\rho
&\!\!\simeq\!\!&
\frac{X^2}{8\,\pi\,\gn^3\,M^2}
\left[
\frac{3\,X}{(1+6\,X)^{1/3}}
+
2
\left(
1
-
\sqrt{1+\frac{3\,X(7\,X-4\,(1+6\,X)^{1/3})}{4\,(1+6\,X)^{2/3}}}
\right)
\right]
\nonumber
\\
&\!\!\simeq\!\!&
\frac{6\,X^3}{8\,\pi\,\gn^3\,M^2}
\ .
\label{rhoM}
\ee
Of course, the pressure is highest in the centre of the object and it is interesting to compare
this central pressure with the density. 
A plot of the two quantities is shown in Fig.~\ref{p0andrho}, where the vertical axis
of the left panel scales in units of $\gn^{-3}\,M^{-2}$.
The variation of the pressure with $\xi$ inside sources of different compactness
is represented in Fig.~\ref{pxi}. 
\par
\begin{figure}[t]
\centering
\includegraphics[width=8cm]{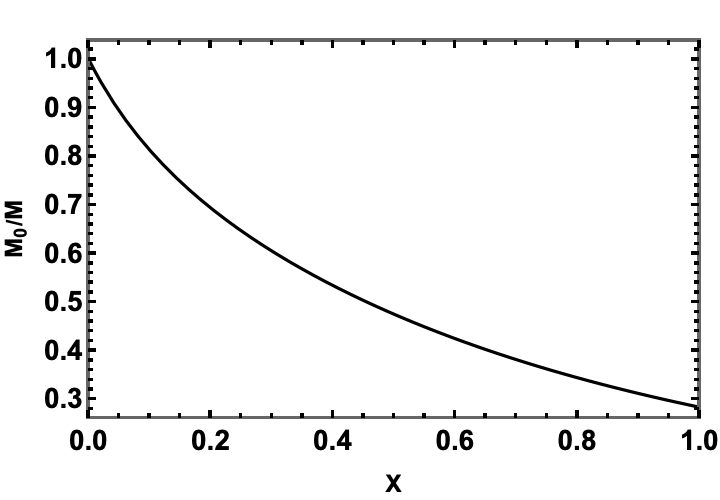}
\caption{Ratio~\eqref{MsM0} of proper mass over ADM-like mass.}
\label{M0M}
\end{figure}

Starting from the two expressions for the density, the one in terms of the proper
mass $M_0$ from Eq.~\eqref{HomDens} and the expression above in terms of
the ADM-like mass $M\simeq M_{\rm s}$, we find the ratio
\be
\frac{M_0}{M_{\rm s}}
&\!\!\simeq\!\!&
\frac{1}{2\,(1+6\,X)^{1/3}}
+
\frac{1}{3\,X}
\left(
1
-
\sqrt{1+\frac{3\,X(7\,X-4\,(1+6\,X)^{1/3})}{4\,(1+6\,X)^{2/3}}}
\right)
\nonumber
\\
&\!\!\simeq\!\!&
1
-
\frac{5}{2}\,X
\ ,
\label{MsM0}
\ee
which is represented graphically in Fig.~\ref{M0M}.
\subsection{Comparisons with numerical solutions}
\label{SS:num1}
Numerical solutions for the potential $V$ and pressure $p$ can be obtained
by solving the system of the second order Eq.~\eqref{EOMV1} and first order Eq.~\eqref{eqp_1}
with the three boundary conditions $V'(\xi=0)=0$ in the centre [see Eq.~\eqref{b0}], $V(\xi=1)=V_R$
[see Eq.~\eqref{VR}] and $p(\xi=1)=0$ at the surface of the star.
Such solutions will depend on the values of $\rho_0$ (hence $M_0$) and $M$.
A solution is then considered acceptable if the fourth boundary condition~\eqref{dVR}
is also satisfied, that is $V'$ is also (sufficiently) continuous across the surface of the star,
which determines $M=M(M_0)$~\cite{Casadio:2019cux,Casadio:2020kbc}.
\par
\begin{figure}
\centering
\includegraphics[width=5.3cm]{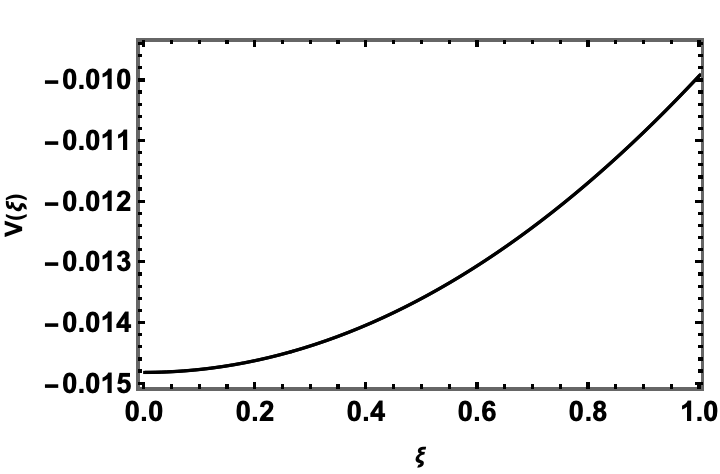}
\includegraphics[width=5.3cm]{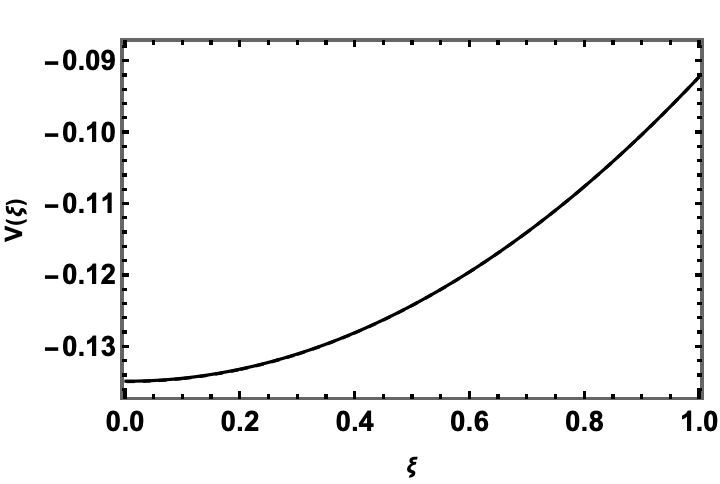}
\includegraphics[width=5.3cm]{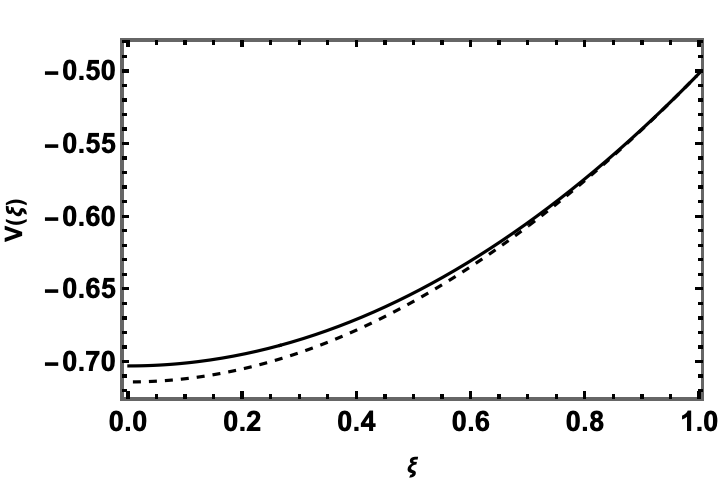}
\\
\includegraphics[width=5.3cm]{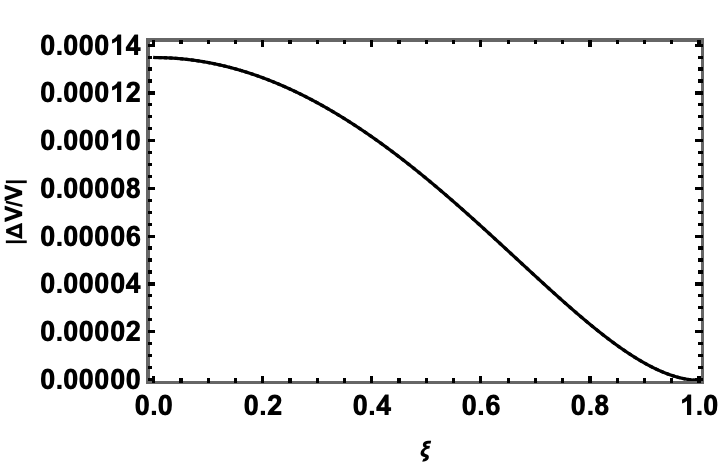}
\includegraphics[width=5.3cm]{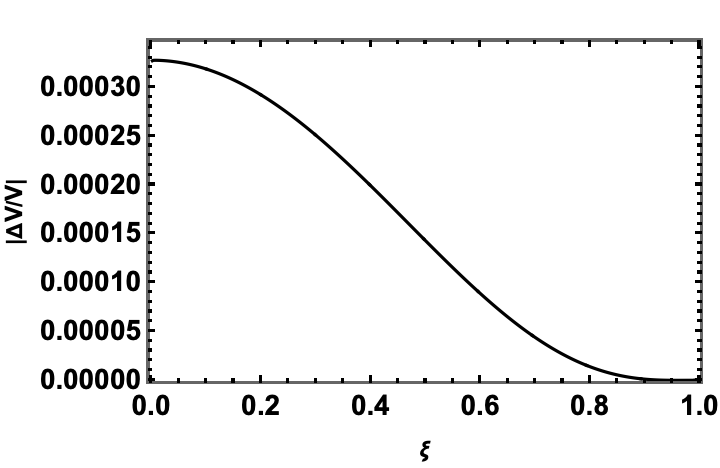}
\includegraphics[width=5.2cm]{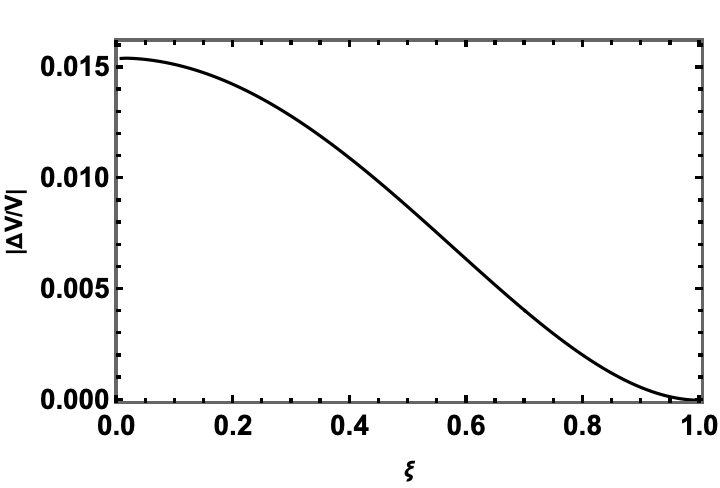}
\caption{Top panels: analytical approximation~\eqref{Vins} (solid lines) and numerical
(dashed lines) BNG potentials for $X=0.01$ (left plot),  $X=0.1$ (center plot) and  $X=0.7$ (right plot).
Bottom panels: relative difference between the numerical and approximate potentials
for the same values of the compactness in the top plots.}
\label{V_errV}
\end{figure}
\begin{figure}
\centering
\includegraphics[width=5.4cm]{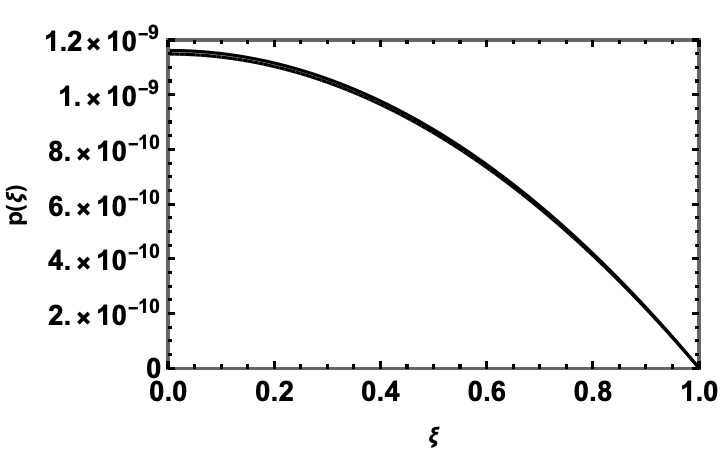}
\includegraphics[width=5.3cm]{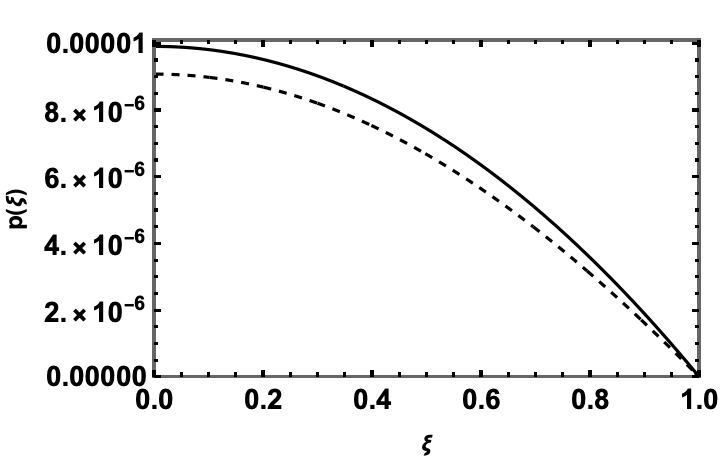}
\includegraphics[width=5.1cm]{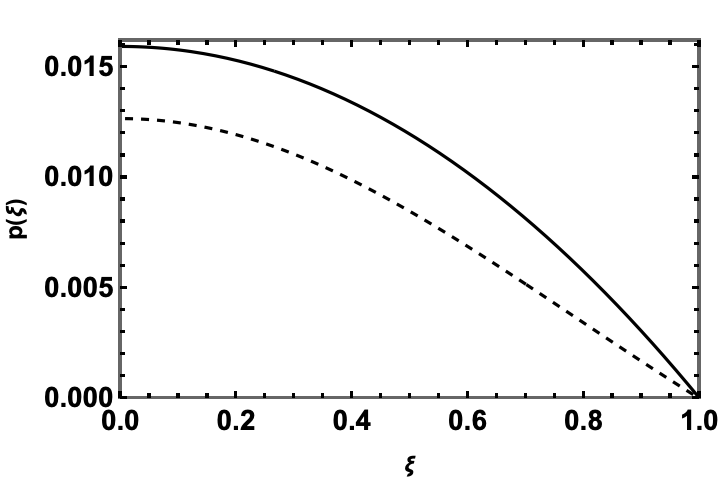}
\\
\includegraphics[width=5.4cm]{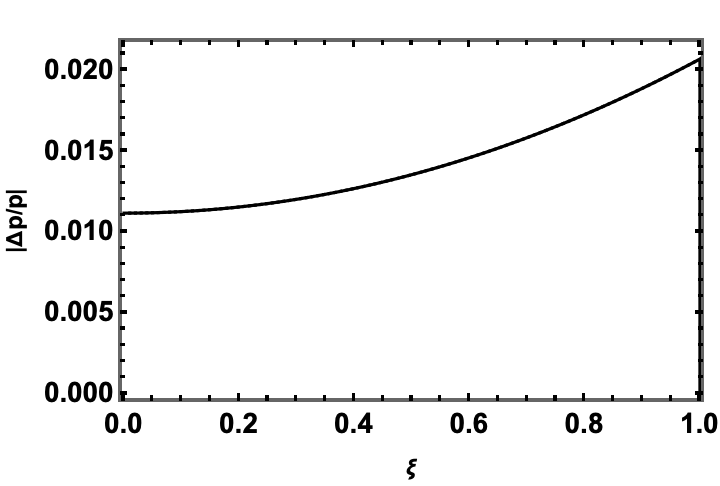}
\includegraphics[width=5.3cm]{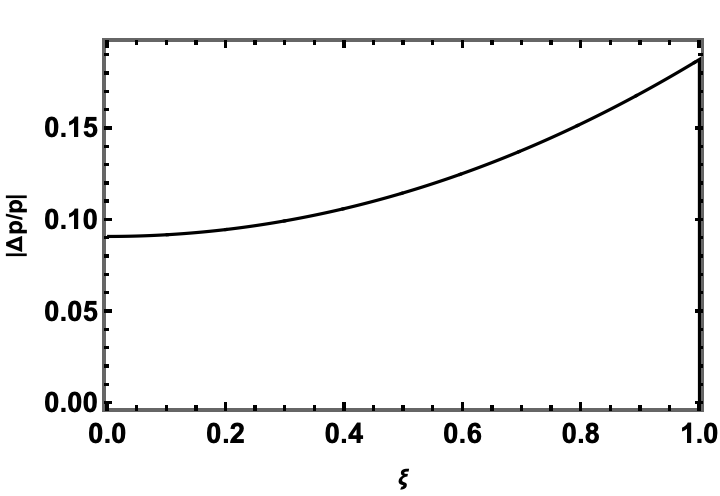}
\includegraphics[width=5.2cm]{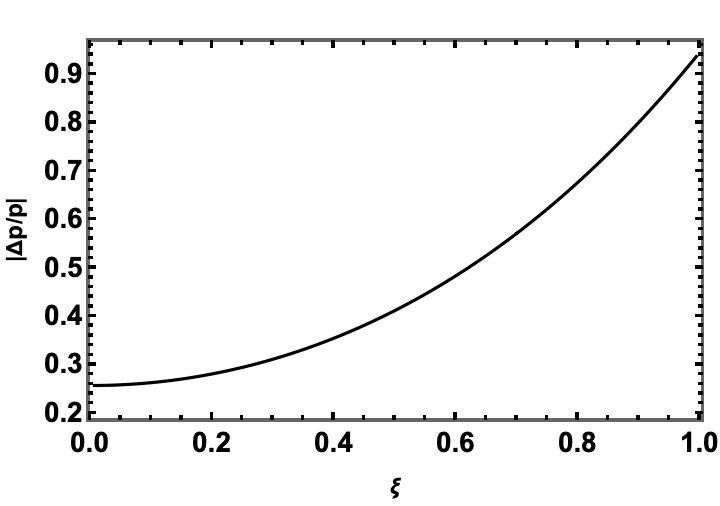}
\caption{Top panels: analytical approximation~\eqref{pinsol} (solid lines) and numerical
(dashed lines) pressures in units of $\gn^{-3}\,M^{-2}$ for $X=0.01$ (left plot),  $X=0.1$ (center plot)
and $X=0.7$ (right plot).
Bottom panels: relative difference between the numerical and approximate pressures
for the same values of the compactness in the top plots.}
\label{p_errp}
\end{figure}

Fig.~\ref{V_errV} presents the comparison between the approximate analytical
solution $V_{\rm s}$ for the potential from Eq.~\eqref{Vins}, and the solution $V_{\rm num}$ obtained
by solving numerically the system of differential Eqs.~\eqref{EOMV1} and~\eqref{eqp_1}
for three different values of the compactness.
Since the difference is very small, the bottom panels of Fig.~\ref{V_errV} show the relative difference
\be
\left|\frac{\Delta V}{V}\right|=
\left|
\frac{V_{\rm num}-V_{\rm s}}{V_{\rm num}}
\right|
\ ,
\label{relV}
\ee
between the two expressions in each of the three cases.  
The comparison between the approximate analytical solution $p_{\rm s}$ in Eq.~\eqref{pinsol}
and the numerically computed pressure $p_{\rm num}$ for the same values of the compactness
is shown in Fig.~\ref{p_errp}, along with their relative difference 
\be
\left|\frac{\Delta p}{p}\right|
=
\left|
\frac{p_{\rm num}-p_{\rm s}}{p_{\rm num}}
\right|
\ .
\label{relp}
\ee
The relative difference between the numerical and approximate analytical solutions
for the potential remains very small for $X\lesssim 0.7$, which shows that the approximation
$V_{\rm s}$ can be used into the (relatively) large compactness regime.
The relative difference for the pressures instead increases to very significant values
already for $X\gtrsim 0.1$.
In particular, the top panels of Fig.~\ref{p_errp} show that $p_{\rm s}$ significantly
overestimates the pressure in the high compactness regime. 
\par
The ratio of the proper mass to ADM-like mass in the three cases discussed above is shown
in Table~\ref{t:M0M}, both for the approximate analytical and numerical cases.
In the high compactness regime, the numerical solutions result in smaller proper mass
to ADM-like masses than the approximate analytical ones. 
\begin{table}[h]
\centering
\begin{tabular}{||c|c|c||}
\hline\hline
$X$ & $\frac{M_0}{M_{\rm s}}$ & $\frac{M_0}{M_{\rm num}}$ \\ \hline
\hline
$0.01$ & $0.976$  & $0.976$  \\ 
\hline
$0.1$ & $0.813$ & $0.807$  \\ 
\hline
$0.7$ & $0.382$ & $0.295$  \\ 
\hline
\hline
\end{tabular}
\caption{Proper mass to ADM-like mass ratios in the approximate analytical case,
respectively in the numerical case for the three compactness values discussed.}
\label{t:M0M}
\end{table}
\subsection{Comparisons with GR}
As long as the isotropic star is larger than the Buchdahl limit~\cite{Buchdahl:1959zz},
below which the GR pressure cannot support a stable configuration,
we can compare the BNG pressure to the GR pressure expressed
in harmonic coordinates at the same order in $\gn$.
We recall that the use of harmonic coordinates is necessary to identify
properly the Newtonian regime and post-Newtonian corrections in GR~\cite{weinberg}.
In particular, we only consider contributions up to second order in $\gn$ for which we
just need the explicit form of the harmonic radial coordinate to first order in $\gn$
[see Appendix~\ref{A:homo}].
In order to compare expressions from the BNG with their GR counterparts,
we consider stars in BNG with values of $M$ equal to the ADM mass $\bar M$ in GR
and the same compactness in both theories, since these are the observables one can
in principle measure.
\par
\begin{figure}
\centering
\includegraphics[width=5.5cm]{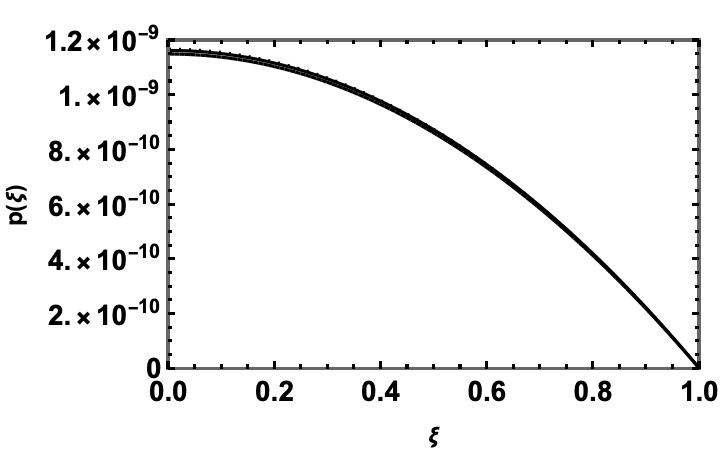}
\includegraphics[width=5.3cm]{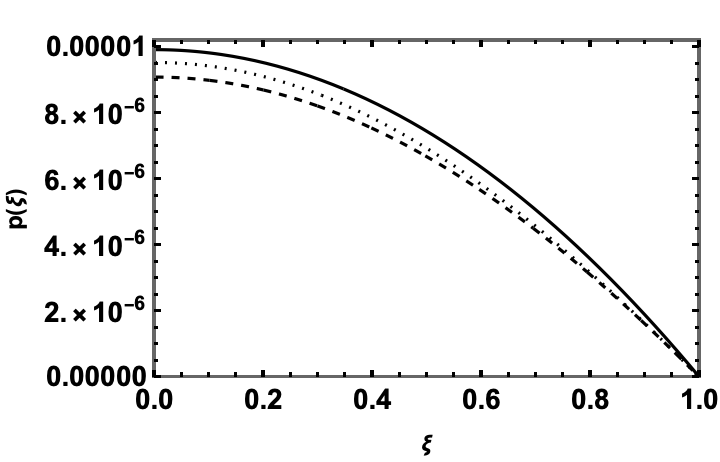}
\includegraphics[width=5.2cm]{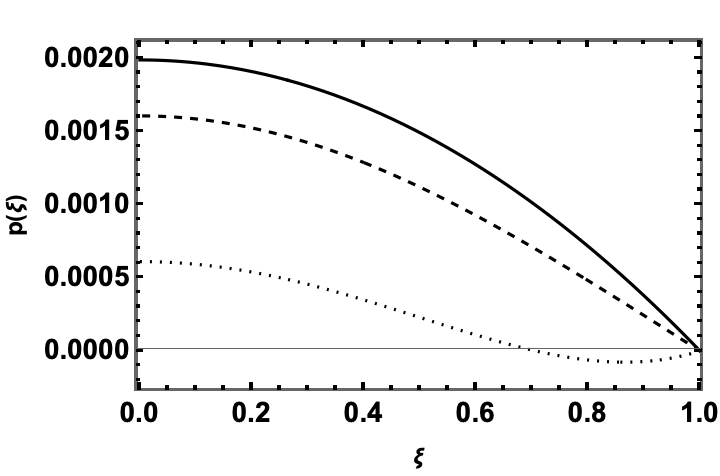}
\caption{Analytic approximation~\eqref{pinsol} for the BNG pressure (solid lines),
numerical BNG pressure (dashed lines) and approximate GR pressure~\eqref{p_harmonic}
(dotted lines) in units of $\gn^{-3}\,M^{-2}$ for $X=0.01$ (left panel), $X=0.1$ (center panel)
and $X=0.4$ (right panel).}
\label{pGR}
\end{figure}

The approximate GR pressure at order $\gn^2$ in harmonic coordinates is taken
from Eq.~\eqref{p_harmonic} and a comparison with the analytic approximation~\eqref{pinsol}, 
respectively numerical solution for the BNG pressure is shown in Fig.~\ref{pGR}.
The approximate GR pressure becomes negative for $X\gtrsim 0.4$, which means that the
expansion to order $\gn^2$ fails at those values of the compactness and the comparison
becomes impossible already before reaching the Buchdahl limit.
However, the approximate BNG and GR pressures are in very good agreement for values of
the compactness $X\lesssim 0.1$.
Above this regime their values  diverge fairly quickly, most likely due to the unreliability
of the GR approximation.  
\par
An important difference between BNG and GR must be emphasised.
As detailed in Appendix~\ref{A:homo}, the Buchdahl limit is reached for $X=4/5$ 
using harmonic coordinates in GR [see Eq.~\eqref{GRB}] and the event horizon
appears at $X=1$, as shown in Eq.~\eqref{GRH}. 
As discussed in Ref.~\cite{Casadio:2019cux}, the location $\rh$ of the horizon
in BNG can be calculated as the radius at which the escape velocity of test particles
equals the speed of light, that is
\be
\label{Vrh}
2\,V(\rh)= -1
\ .
\ee
When the compactness is small, $V(0)>-1/2$ and no horizon exists.
For $X\simeq 0.46$ one has $V(0)=-1/2$ and a horizon appears in the centre.
For larger values of the compactness, the potential well deepens and the size
$\rh$ of the horizon increases to equal the radius $R$ of the source for
$X \simeq 0.69$, when the star becomes a BNG black hole. 
The above reasoning is what motivated our choice of compactness values 
shown in the plot in Fig.~\ref{Fig1}.
For even larger compactness, the horizon radius is located in the outer vacuum
given in Eq.~\eqref{sol0}. 
An important point to note is that BNG objects described this far do not exhibit
an equivalent of the Buchdahl limit~\cite{Casadio:2019cux,Casadio:2020kbc}. 
This implies that both pressure and density remain well-behaved functions
as the compactness increases, up until the object forms a black hole. 
A more detailed discussion of the phenomenological implications resulting 
from the absence of a Buchdahl limit is presented in the concluding Section~\ref{S:Disc}, 
since it is a more general topic that not only applies to the conservation equation 
used in this section, but to the forms investigated in previous works as well. 
 \subsection{Gravitational energy}
We can calculate the gravitational potential energy $U_{\rm G}$ starting from the effective
Hamiltonian $H[V]=-L[V]$, with the Lagrangian given by Eq.~\eqref{LagrV}
(with all couplings set to one).
The gravitational potential energy can be written as the sum of two terms,
with the second term also containing two parts~\cite{Casadio:2019cux}, namely 
\be
U_{\rm G} 
&=&
U_{\rm BG}
+
U_{\rm GG}
\nonumber
\\
&=&
U_{\rm BG}
+
U_{\rm GG}^{\rm in} 
+
U_{\rm GG}^{\rm out} 
\nonumber
\\
&=&
4\,\pi
\int_0^\infty
r^2\,\d r
\left(\rho+p\right) V_{\rm in}\left(1-2\,V_{\rm in}\right)
\nonumber
\\
&&
+
\frac{1}{2\,\gn}
\int_0^R
r^2\,\d r\,
\left(V_{\rm in}'\right)^2
\left(1-4\,V_{\rm in}\right) 
\nonumber
\\
&&+
\frac{1}{2\,\gn}
\int_R^\infty
r^2\,\d r\,
\left(V_{\rm out}'\right)^2
\left(1-4\,V_{\rm out}\right)
\ .
\label{UG}
\ee
The result is very cumbersome, but the total gravitational energy is plotted as a function of 
the compactness in Fig.~\ref{UG}.
\par
In the low compactness limit $X\ll 1$, and keeping the first two lowest order terms,
one obtains
\be
U_{\rm G} 
\simeq
-\frac{3\, \gn\, M^2}{5\,R}
+
\frac{9\, \gn^2\, M^3}{7\,R^2}\ ,
\label{UGlow}
\ee
where the compactness $X$ was written explicitly in terms of the ADM-like mass and the radius
of the star.
We immediately notice that the lowest order term is indeed the usual Newtonian gravitational energy
for a homogenous star.
\begin{figure}[tp]
\centering
\includegraphics[width=8cm]{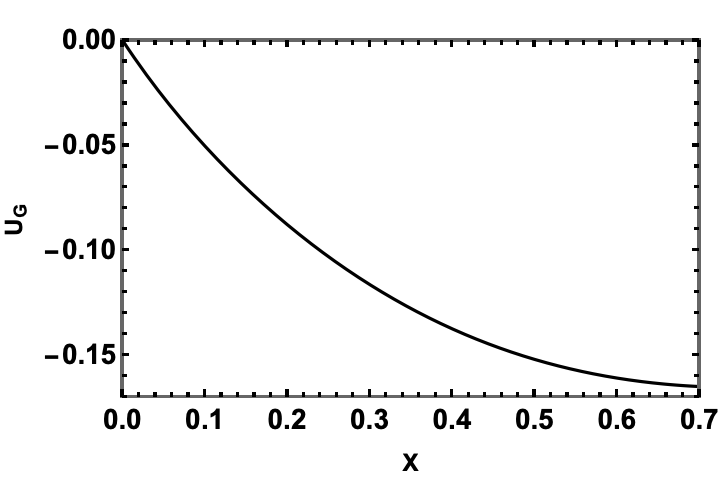}
\caption{Gravitational energy $U_{\rm G}$ in units of ADM-like mass $M$.}
\label{UG}
\end{figure}
\section{Additional pressure and potential terms}
\setcounter{equation}{0}
\label{S:addPV}
We now consider the full conservation equation~\eqref{eqp}, that is
\be
p'
\simeq
-
\left( \rho + p\right) V' - 4\,\pi\,\gn\,r\, p\,\rho -2\,\rho\,V\,V'
\ .
\label{eqp_2}
\ee 
The difficulty of obtaining analytic solutions is similar to the case discussed in
Section~\ref{S:addP} but results in much more cumbersome expressions for the interior
pressure and density, as well as relationship between the proper and ADM-like mass. 
\par
The analytic approximation for the potential obtained by Taylor expanding around $r=0$
is fully constrained by imposing the boundary conditions~\eqref{VR}, \eqref{dVR} and
\eqref{b0} and remains the same $V_{\rm s}$ given in Eq.~\eqref{Vins}.
The differences within this approximation are therefore seen in the expressions for the
pressure and density. The full expressions are very cumbersome to display, but their first two 
lowest order terms in the small compactness limit $X\ll 1$ are given by
\be
p_{\rm s}
\simeq
\frac{3\,X^4\left(2 - 11\,X\right)\left(1 - \xi^2\right)}{16\,\pi\,\gn^3\,M^2} \ ,
\label{pin_v1}
\ee
for the pressure and 
\be
\rho
\simeq
\frac{3\,X^3\left(2 - 5\,X\right)}{8\,\pi\,\gn^3\,M^2}
\label{rho_v1}
\ee
for the density.
\begin{figure}[tp]
\centering
\includegraphics[width=5.3cm]{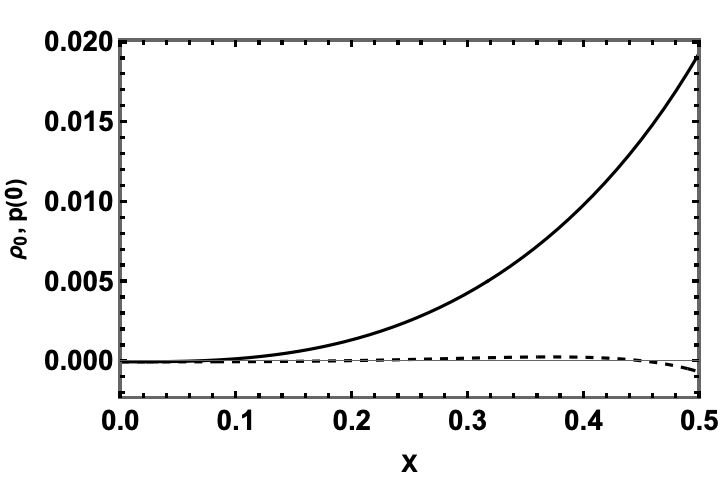}
\includegraphics[width=5.3cm]{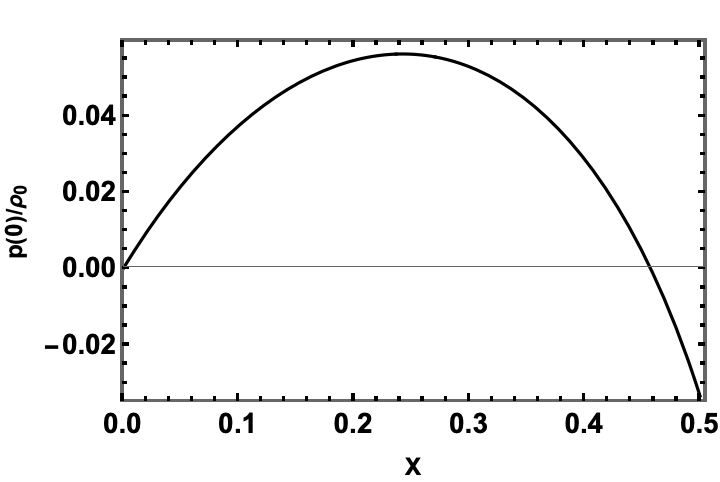}
\includegraphics[width=5.5cm]{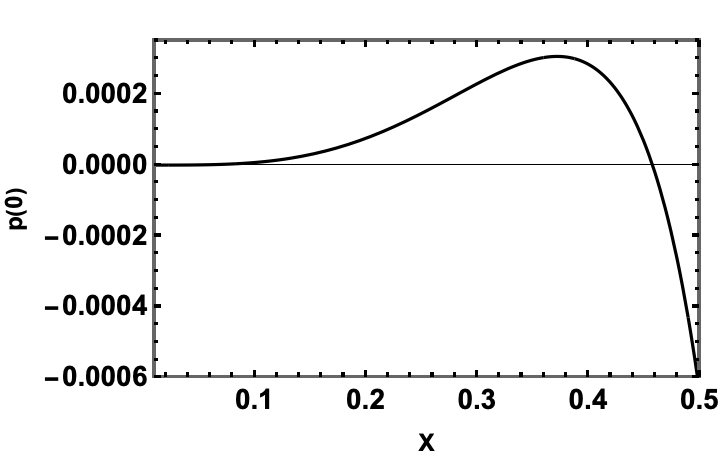}
\caption{Left panel: density (solid curve) and central pressure (dashed curve) (in units of $\gn^{-3}\,M^{-2}$).
Center panel: ratio of the central pressure and density.
Right panel: central pressure.}
\label{p0andrho_v1}
\end{figure}
\par
\begin{figure}[]
\centering
\includegraphics[width=8cm]{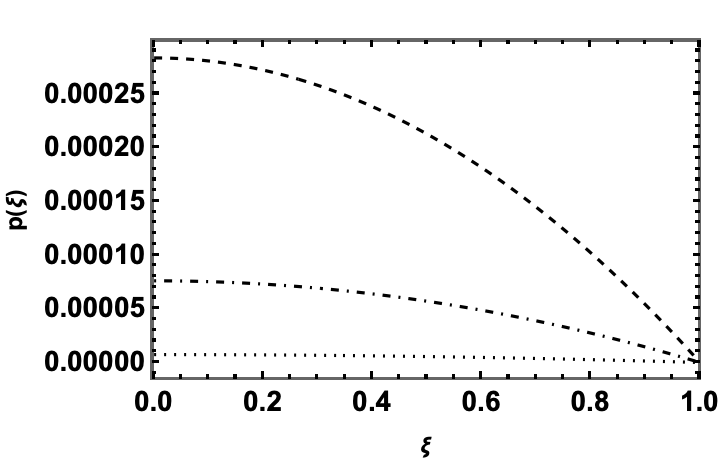}
\caption{Pressure for compactness $X=0.1$ (dotted), $X=0.2$ (dot-dashed) and $X=0.4$ (dashed) in
units of $\gn^{-3}\,M^{-2}$.}
\label{pxi_v1}
\end{figure}
\begin{figure}[tp]
\centering
\includegraphics[width=8cm]{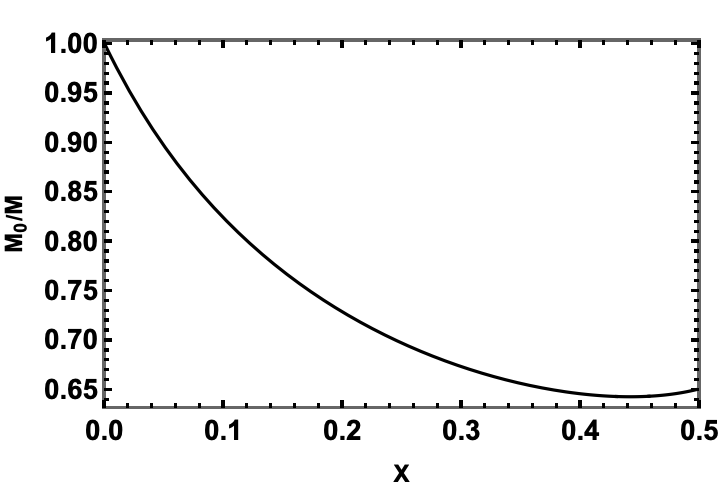}
\caption{Ratio between proper and ADM-like mass.}
\label{M0M_v1}
\end{figure}

Fig.~\ref{p0andrho_v1} shows the plots of the density and central pressure, the ratio of the central pressure
to the density, and the central pressure, all as functions of the compactness.
These plots show that the additional term from the conservation equation results in considerable
differences with respect to the previous case.
First of all, in contrast to the previous case, while the density is still increasing, the pressure
remains small and it eventually goes through zero and becomes negative somewhere in the range
$0.4<X<0.5$.
This is also very different from the GR case in which the central pressure diverges
at the Buchdahl limit.
Another detail to remark is that the ratio of the central pressure to the density goes through
a maximum and then decreases, only to become negative as the pressure becomes negative. 
Fig.~\ref{pxi_v1} shows some plots for the pressure for values of the compactness in the range
in which the pressure is well behaved.
In the same range, the ratio of $M_0/M$ decreases for increasing compactness as expected (Fig.~\ref{M0M_v1}).
It is worth emphasising that something unexpected happens well before the object reaches
the compactness of a black hole:
the central pressure becomes negative before the compactness reaches
$X\simeq 0.46$, the value for a horizon to appear in the centre of the object. 
\subsection{Comparisons with numerical solutions and GR}
\label{SS:num2}
Numerical solutions can be obtained from the same algorithm described at the beginning of
Section~\ref{SS:num1}.
Fig.~\ref{V_errV_v1} shows the comparison between the approximate analytic solution for the potential 
$V_{\rm s}$ and the numerical solution $V_{\rm num}$, along with the corresponding relative
difference for each compactness value.
The plots show that the approximate analytic solution remains very good for $X\lesssim 0.5$
(down from $X\lesssim 0.7$ in the previous case). 
\par
The behaviour of the pressure at relatively high compactness is again more interesting. 
As noted earlier, the analytic approximation becomes negative somewhere in the interval $0.4<X<0.5$. 
The numerical simulations also lead to negative values of the pressure, but the sign change
occurs somewhere in the range $0.5<X<0.6$.
This is the reason why the largest compactness value shown in Fig.~\ref{V_errV_v1} is $X= 0.5$.
\par
We recall from Fig.~\ref{pGR} that the approximate GR pressure becomes negative for $X\gtrsim 0.4$.
Fig.~\ref{pGR_v1} now shows a comparison between the analytical approximation $p_{\rm s}$
and numerical BNG pressure $p_{\rm num}$, along with the approximate GR solution. 
These three quantities remain very close In the small compactness regime $X\ll 1$.
For $X\simeq 0.1$ the numerical pressure matches the analytic approximation and they both
predict a smaller pressure than the GR approximation. 
For $X\simeq 0.5$, the numerical pressure is positive throughout the entire star volume,
while the other two approximations are negative. 
As the compactness further increases, the numerical solution for the system of differential
equations results in negative pressure somewhere in the interval $0.5<X<0.6$.
To summarise, while the numerical solutions for the BNG potential are similar to those obtained
in the previous section all the way up to compactness $X\simeq 0.7$
(the potential for this value is not displayed herein), both solutions for the pressure become
negative before the compactness of the star is large enough for it to become a black hole.  
{\color{blue}
\begin{figure}
\centering
\includegraphics[width=5.4cm]{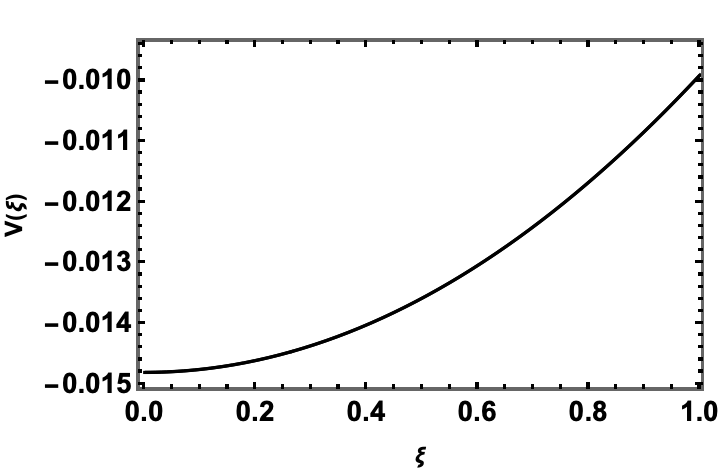}
\includegraphics[width=5.3cm]{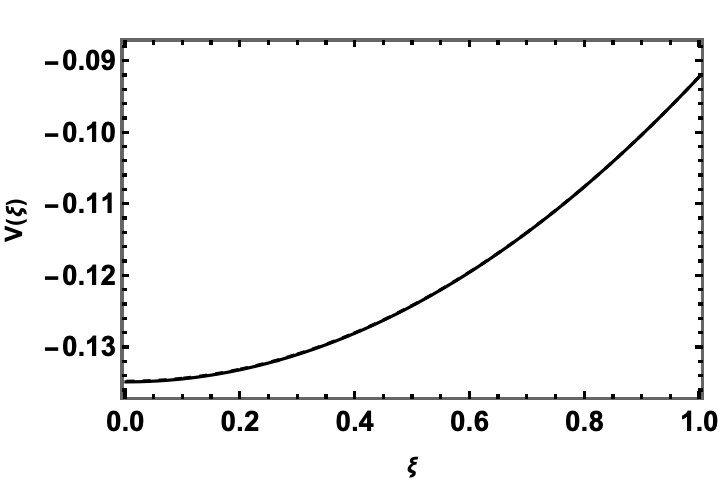}
\includegraphics[width=5.2cm]{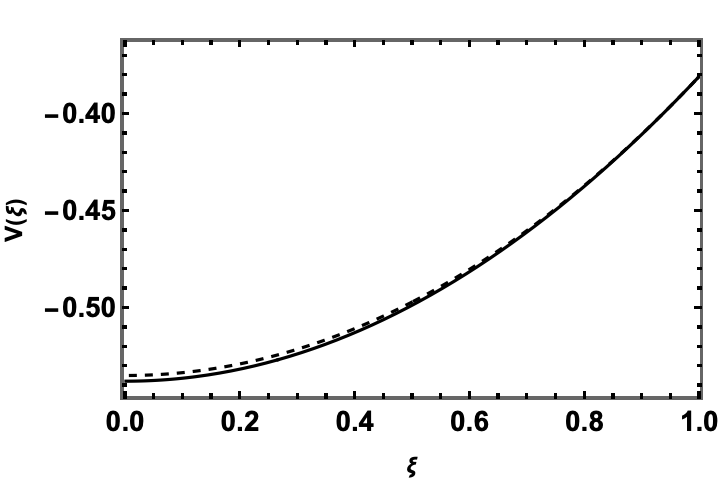}
\\
\includegraphics[width=5.3cm]{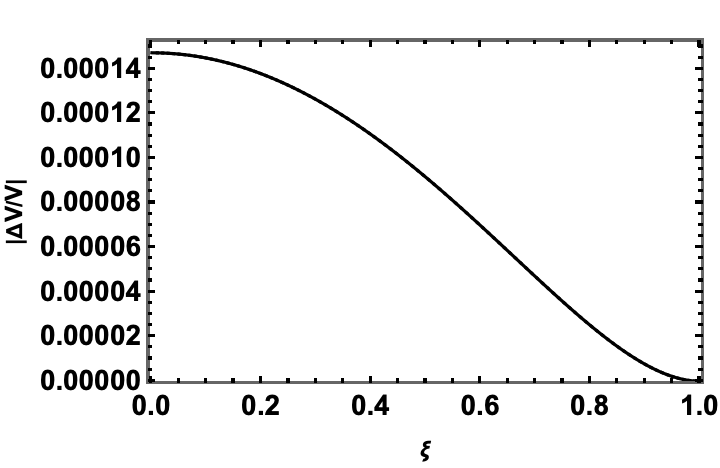}
\includegraphics[width=5.3cm]{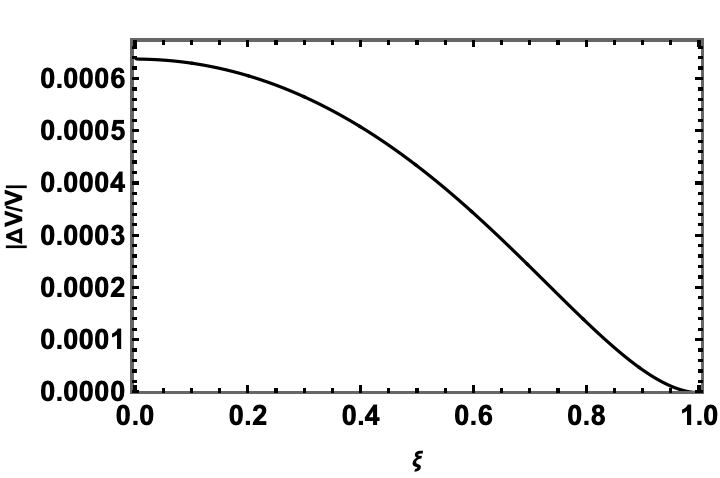}
\includegraphics[width=5.2cm]{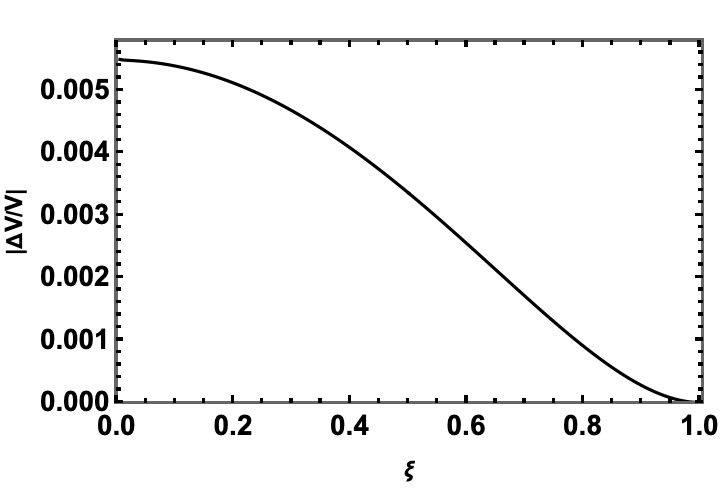}
\caption{Top panels: analytical approximation (solid lines) and numerical (dashed lines) 
BNG potential for $X=0.01$ (left plot),  $X=0.1$ (center plot) and  $X=0.5$ (right plot).
Bottom panels: relative difference between the numerical and approximate potentials
for the same values of the compactness.}
\label{V_errV_v1}
\end{figure}
\begin{figure}
\centering
\includegraphics[width=5.5cm]{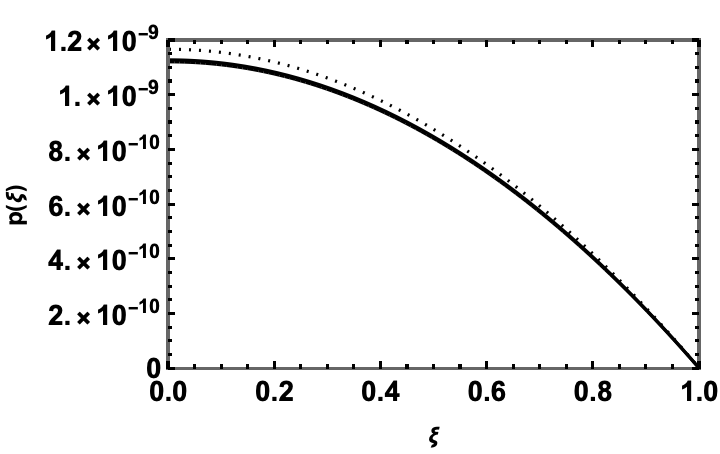}
\includegraphics[width=5.3cm]{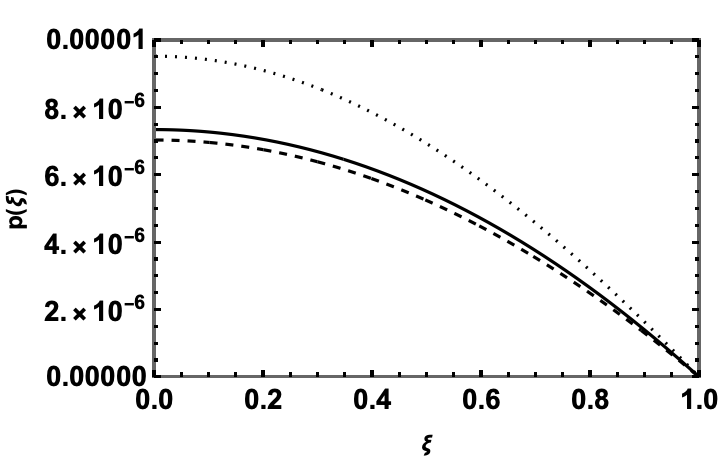}
\includegraphics[width=5.2cm]{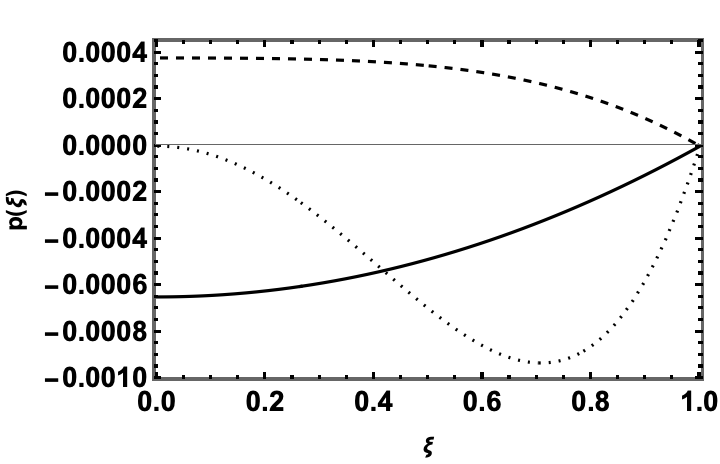}
\caption{Analytic approximation for the BNG pressure (solid lines), numerical BNG pressure (dashed lines)
and approximate GR pressure from Eq.~\eqref{p_harmonic} (dotted lines) in units of $\gn^{-3}\,M^{-2}$
for $X=0.01$ (left panel), $X=0.1$ (center panel) and $X=0.5$ (right panel).}
\label{pGR_v1}
\end{figure}
}
\section{Discussion}
\setcounter{equation}{0}
\label{S:Disc}
The results obtained in Section~\ref{S:addP} show that, within the discussed approximation,
the conservation equation with the additional pressure term included allows the formation
of sufficiently compact objects to be enclosed by a horizon.
When compared to numerical solutions, the approximate solution for the BNG potential
closely follows the numerical result throughout the entire range taken into consideration.
This can also be inferred from the plots that display the relative difference between the two.
The approximate analytical solution for the pressure is larger than the one obtained by solving
the equations numerically.
Their relative difference is largest when the compactness is $X\simeq 0.7$. 
\par
The BNG pressure and the approximate GR pressure were also compared and the three
quantities follow one another closely all the way up to $X\simeq 0.1$.
Beyond this value of the compactness, the GR approximation diverges fairly quickly, likely because the
approximation (obtained for $X\ll 1$) becomes problematic in the high compactness regime.
Notably, the GR pressure becomes negative already around $X\simeq 0.4$. 
\par
It is worth mentioning that while the BNG horizon appears at $X\simeq0.69$,
the GR horizon forms when $X=1$.
The BNG compactness for which the horizon is as large as the star
is calculated using the exact BNG potential in vacuum, where no approximation is needed.
In contrast to GR, BNG objets do not show an equivalent to the Buchdahl limit,
which means that both the pressure and the density can be well behaved functions
as the compactness increases until the object becomes a black hole. 
Finally, while the expressions for the different components of the gravitational energy
are cumbersome, the small-compactness limit recovers the familiar Newtonian
gravitational energy at leading order.
\par
In Section~\ref{S:addPV}, an additional term proportional to the potential and its first derivative
was included in the conservation equation, leading to deviations from previous results. 
While the match between the approximate analytic and numeric potential remains very good
up to the fairly large value of $X\simeq 0.4$, the extra term affects the analytic approximation
for the pressure significantly, causing it to remain much smaller relative to the density,
as illustrated in the middle panel of Fig.~\ref{p0andrho_v1}.
The right panel of the same figure also highlights the most critical difference between the two cases:
the pressure starts to decrease for $X \geq 0.4$ and quickly becomes negative
before the formation of a horizon.
When the pressure is evaluated numerically, this transition to negative pressure takes
place somewhere in the range $0.5<X<0.6$, before the compactness is large enough
for the object to be a black hole.
\par 
What both cases analysed here show is that the analytic approximation $V_{\rm s}$
for the BNG potential in Eq.~\eqref{Vins} works very well within the source for all values of the
compactness up to the appearance of a horizon at $X\sim 0.7$.
Quite interestingly, this is approximately the same value at which the second order
GR expansion of the pressure fails in the centre of the star [see Appendix~\ref{A:homo}].
\par
The absence of a Buchdahl limit represents a fundamental difference between stars in BNG
and their description in GR.
The second essential difference is the existence of non-singular pressure and density profiles
that can be hidden behind the event horizon.
While the scientific community currently lacks the necessary knowledge or means to look behind
the horizon and investigate the interior of black holes, if gaining such insights is even possible,
searching for a Buchdahl limit (or its absence) in ultra-dense stars on the verge of becoming
black holes may provide an indirect way to understand these extreme conditions. 
Future experiments, similar to the Event Horizon Telescope~\cite{EventHorizonTelescope:2022wkp},
may eventually achieve the precision needed to observe such dense stars and measure their
compactness in this regime.
If objects that violate the Buchdahl limit are discovered, this could also suggest the possible
existence of non-singular density profiles for objects hidden behind the horizon.
\section*{Acknowledgments}
R.C.~is partially supported by the INFN grant FLAG and his work has also been carried out in
the framework of activities of the National Group of Mathematical Physics (GNFM, INdAM).
O.M.~was supported by the Romanian Ministry of Research, Innovation and Digitalization
under the Romanian National Core Program LAPLAS VII - contract no. 30N/2023.
\appendix 
\section{Post-Newtonian TOV equation}
\setcounter{equation}{0}
\label{A:NpN}
The general condition of equilibrium for self-gravitating spherically symmetric and isotropic
compact objects in GR is given by the TOV equation~\cite{Tolman:1939jz,Oppenheimer:1939ne},
which can be easily derived starting from a static and spherically symmetric metric in the form
\be
\d s^2 =
-e^{\nu}\,\d t^2
+
e^{\lambda}\,\d \bar r^2
+\bar r^2\,\d\Omega^2
\ ,
\label{staticg}
\ee
where $\bar r$ is the areal radius, $\lambda=\lambda(\bar r)$ and $\nu=\nu(\bar r)$.
Assuming matter is given by a static perfect fluid with proper energy density $\rho=\rho(\bar r)$
and isotropic hydrostatic pressure $p=p(\bar r)$, the Einstein equations read~\cite{Stephani:2004ud} 
\be
G^0_{\ 0}
&\!\!=\!\!&
-e^{-\lambda}
\left(\frac{\lambda'}{\bar r}-\frac{1}{\bar r^2}\right)
+
\frac{1}{\bar r^2}
=
-8\,\pi\,\gn\,\rho
\label{G00}
\\
\nonumber
\\
G^1_{\ 1}
&\!\!=\!\!&
e^{-\lambda}
\left(\frac{\nu'}{\bar r}+\frac{1}{\bar r^2}\right)
+\frac{1}{\bar r^2}
=
8\,\pi\,\gn\,p
\label{G11}
\\
\nonumber
\\
G^2_{\ 2} 
&\!\!=\!\!&
G^3_{\ 3}
=
e^{-\lambda}
\left[\frac{\nu''}{2}+\frac{(\nu')^2}{4}-\frac{\nu'\,\lambda'}{4}
+\frac{\nu'-\lambda'}{2\,\bar r}
\right]
=
8\,\pi\,\gn\,p
\label{G22}
\ ,
\ee
where primes denote derivatives with respect to $\bar r$.
From Eq.~\eqref{G00}, one then obtains 
\be
e^{-\lambda}
=
1-\frac{2\,\gn\,m(\bar r)}{\bar r}
\ ,
\label{m(r)}
\ee
where the Misner-Sharp-Hernandez mass function~\cite{Misner:1964je,Hernandez:1966zia}
\be
m(\bar r)
=
{4\,\pi} \int_0^{\bar r} \rho(x)\,x^2\,\d x
\label{massclass}
\ee
is such that the total mass of a star of radius $\bar r=\bar R$ is given by $\bar M=m(\bar R)$,
which is the Arnowitt-Deser-Misner~(ADM) mass~\cite{Arnowitt:1959ah}
of the system in the outer Schwarzschild metric~\cite{Schwarzschild:1916uq}.
The Bianchi identity $\nabla_\mu G^\mu_{\ \nu}=0$ then implies the conservation equation
$\nabla_\mu T^\mu_{\ \nu}=0$ or
\be
\nabla_\mu T^\mu_{\ 1}
=
p'
+
\frac{\nu'}{2}\left(\rho+p\right)
=
0
\ .
\label{contR}
\ee
The TOV equation is now obtained by solving for $\nu'$ from Eq.~\eqref{G11}
and replacing its expression into Eq.~\eqref{contR}, 
\be
p'
=
-\left(\rho+\epsilon \,p\right)
\frac{\gn}{\bar r^2}\left[m+4\,\pi\, \epsilon \, \bar r^3\,p\right]
\left[1-\frac{2\,\gn\,m}{\bar r}\right]^{-1}
\ ,
\label{grTOV1}
\ee
where $\epsilon \simeq 1/c^2$ (with $\epsilon \sim \gn \ll 1$) was introduced
to keep track of the factors inversely proportional to the speed of light.
Let us also replace the GR expression proportional to the Misner-Sharp-Hernandez
mass with an effective potential
\be
\frac{\gn\,m}{\bar r}
\to
V
\equiv
\gn\,\mathcal{V} +\gn^2 \,\mathcal{W}
\ .
\ee
By expanding Eq.~\eqref{grTOV1} up to second order (in $\epsilon$ and $\gn$), we then find
\be
p'
\simeq
-
\gn\,\rho\, \mathcal{V}'
-
\epsilon\,\gn\,p\left(\mathcal{V}' +4\,\pi\,\bar r\,\rho \right)
-
\gn^2\,\rho
\left(
\mathcal{W}'
+
2\,\mathcal{V}\,\mathcal{V}'
\right)
\ ,
\label{TOVmod2}
\ee
whose leading order clearly reproduces the Newtonian condition~\eqref{eqP_OLD} with
$V=\gn\,\mathcal V$ and $\bar r=r$~\cite{weinberg}.
\par
Going beyond leading order is of course ambiguous, since $\gn$ and $\epsilon$ are not
dimensionless and the validity of the above (formal) expansion can only be checked
{\em a posteriori\/} by estimating the size of the neglected higher-order terms for
any given solution.
In particular, we rewrite Eq.~\eqref{TOVmod2} as
\be
p'
\simeq
-
\left( \rho + \epsilon\, p\right) V' - 4\,\pi\,\gn\,\epsilon\,\bar r\, p\,\rho-2\,\rho\,V\,V'
\ ,
\label{TOVmod2a}
\ee
and test the effects of the various terms on a homogeneous star, for which the
exact Newtonian and GR solutions are recalled in Appendix~\ref{A:homo}.
\section{Homogenous stars in GR and Newtonian approximation}
\setcounter{equation}{0}
\label{A:homo}
Let us consider a ball of radius $\bar r=\bar R$ and homogenous density~\eqref{HomDens}.
In GR, the exact solution of the corresponding Einstein equation is given by
the Schwarzschild metrics, namely Eq.~\eqref{staticg} with~\cite{Schwarzschild:1916uq,Schwarzschild:1916ae}
\be
e^{-\lambda}
=
\left\{
\begin{array}{ll}
1
-
\strut\displaystyle\frac{2\,\gn\,\bar M\,\bar r^2}{\bar R}
\qquad
&
{\rm for}\ 0\le \bar r< \bar R
\\
\\
1-\strut\displaystyle\frac{2\,\gn\,\bar M}{\bar r}
&
{\rm for}\ \bar R<\bar r
\end{array}
\right.
\ee
and
\be
e^\nu
=
\left\{
\begin{array}{ll}
\strut\displaystyle\frac{1}{4}
\left(
3\,\sqrt{1-\frac{2\,\gn\,\bar M}{\bar R}}
-
\sqrt{1-\frac{2\,\gn\,\bar M\,\bar r^2}{\bar R^3}}
\right)^2
\qquad
&
{\rm for}\ 0\le \bar r< \bar R
\\
\\
1-\strut\displaystyle\frac{2\,\gn\,\bar M}{\bar r}
&
{\rm for}\ \bar R<\bar r
\ ,
\end{array}
\right.
\ee
where $\bar R$ is the areal radius of the star and $\bar M$ its ADM mass.
\par
The weak-field expansion of the exact solution is obtained by formally~\footnote{It is in fact 
an expansion in the compactness $X=\gn\,\bar M/\bar R$, as can be more easily seen by introducing the
dimensionless areal radius $\bar r/\bar R$.}
expanding in $\gn$ and yields
\be
V_{\rm GR}
=
\frac{e^\nu-1}{2}
=
V_{\rm N}
+
\mathcal{O}(\gn^2)
\ ,
\label{Vgr1}
\ee
where the Newtonian solution is given by
\be
V_{\rm N}
=
\left\{
\begin{array}{ll}
\strut\displaystyle\frac{\gn\,M_0}{2\,R^3}
\left(r^2-3\,R^2\right)
\qquad
&
{\rm for}\ 0\le r< R
\\
\\
\strut\displaystyle-\frac{\gn\,M_0}{r}
&
{\rm for}\ R<r
\ ,
\end{array}
\right.
\ee
in which we set $\bar r=r$, $\bar R=R$ and $\bar M=M_0$, since 
$e^\nu\sim e^\lambda\sim 1+\mathcal{O}(\gn)$ implies that
$\bar r=r+\mathcal{O}(\gn)$.
The corresponding exact GR pressure for $0\le \bar r< \bar R$ is given by
\be
p
=
\rho_0\,
\frac{\sqrt{1-{2\,\gn\,\bar M}/{\bar R}}-\sqrt{1-{2\,\gn\,\bar M\,\bar r^2}/{\bar R^3}}}
{\sqrt{1-{2\,\gn\,\bar M\,\bar r^2}/{\bar R^3}}-3\,\sqrt{1-{2\,\gn\,\bar M}/{\bar R}}}
\ ,
\label{pschw}
\ee
which diverges in the Buchdahl limit $\bar R\to \bar R_{\rm B}=(9/4)\,\gn\,\bar M$~\cite{Buchdahl:1959zz}.
At first order in $\gn$, the above pressure reads
\be
p
\simeq
\left(
1
-
\frac{r^2}{R^2}
\right)
\frac{3\,\gn\,M_0^2}{8\,\pi\,R^4}
\ ,
\ee
and is therefore of the same order as $V_{\rm N}$.
In order to neglect the pressure, one must therefore also expand for $c\to\infty$,
as recalled in Appendix~\ref{A:NpN}.
\par
In order to properly determine terms of higher order in $\gn$ from the exact GR expressions,
however, one needs to employ the harmonic coordinate $r=r(\bar  r)$, which must
satisfy~\cite{weinberg,Casadio:2021gdf}
\be
\frac{\d}{\d \bar r}
\left[
\bar r^2\,e^{(\nu-\lambda)/2}\,\frac{\d r}{\d \bar r}
\right]
=
2\,e^{(\nu+\lambda)/2}\,r
\ .
\label{eqH}
\ee
Since $V\sim \mathcal{O}(\gn)$, the explicit knowledge of $r=r(\bar r)$
becomes necessary only from the second order in $\gn$ as anticipated.
\par
For $\bar R<\bar r$, Eq.~\eqref{eqH} reads
\be
\frac{\d}{\d\bar  r}
\left[
\bar r^2\left(1-\frac{2\,\gn\,\bar M}{\bar r}\right)
\frac{\d r}{\d\bar  r}
\right]
=
2\,r
\ee
and one finds
\be
r
=
\bar r-\gn\,\bar M
\ ,
\ee
which yields the post-Newtonian expansion
\be
V_{\rm GR}
\simeq
V_{\rm N}
+
\frac{\gn^2\,\bar M^2}{\bar r^2}
\ .
\label{Vgr2}
\ee
\par
For $0\le \bar r< \bar R$, however, Eq.~\eqref{eqH} becomes much more involved,
\be
&&
\frac{\d}{\d\bar  r}
\left[
\frac{\bar r^2}{2}\,\sqrt{1-\frac{2\,\gn\,\bar M\,\bar r^2}{\bar R^3}}
\left(
3\,\sqrt{1-\frac{2\,\gn\,\bar M}{\bar R}}
-
\sqrt{1-\frac{2\,\gn\,\bar M\,\bar r^2}{\bar R^3}}
\right)
\frac{\d r}{\d\bar  r}
\right]
\nonumber
\\
&&
=
\frac{
3\,\sqrt{1-{2\,\gn\,\bar M}/{\bar R}}
-
\sqrt{1-{2\,\gn\,\bar M\,\bar r^2}/{\bar R^3}}}
{\sqrt{1-{2\,\gn\,\bar M\,\bar r^2}/{\bar R^3}}}\, r
\ .
\label{harmonicR}
\ee
Since we only wish to include terms up to the second order in $\gn$ which arise as corrections
to the first order Newtonian expressions, it is sufficient to compute solutions of Eq.~\eqref{harmonicR}
to first order in $\gn$.
Eq.~\eqref{harmonicR} to order $\gn$ reads
\be
\frac{\d}{\d\bar  r}
\left\{
\bar r^2
\left[1
-
\left(3+\frac{\bar r^2}{\bar R^2}\right)
\frac{\gn\,\bar M}{2\,\bar R}
\right]
\frac{\d r}{\d \bar r}
\right\}
\simeq
2
\left[
1
-
\left(
1-\frac{\bar r^2}{\bar R^2}
\right)
\frac{3\,\gn\,\bar M}{\bar R}
\right]
r
\ .
\ee
Assuming $r=\bar r+\gn\,f(\bar r)$, we then obtain
\be
\bar r^2\,f''+2\,\bar r\,f'-2\,f
=
\frac{5\,\bar M\,\bar r^3}{\bar R^3}
\ ,
\ee
whose general solution is given by
\be
f
=
a\,\bar r
+
\frac{b}{\bar r}
+
\frac{\bar M\,\bar r^3}{2\,\bar R^3}
\ .
\ee
We must set the integration constant $b=0$ in order for $r$ to be well-defined for $\bar r=0$.
The constant $a$ can then be so chosen to ensure continuity of $r$ across $\bar r=\bar R$,
that is $f(\bar R)=-\bar M$, which yields
\be
a
=
-\frac{3\,\bar M}{2\,\bar R}
\ .
\ee
For $0\le \bar r< \bar R$, we thus have
\be
r
\simeq
\bar r
\left(
1
-
\frac{3\,\gn\,\bar M}{2\,\bar R}
\right)
+
\frac{\gn\,\bar M\,\bar r^3}{2\,\bar R^3}
\ ,
\label{trr}
\ee
with $r(0)=0$ and $R\equiv r(\bar R)=\bar R-\gn\,\bar M$.
From this expression one finds that the horizon radius in harmonic coordinates is located at
\be
r(\bar R_{\rm H})
=
\gn\,\bar M
\quad
\Rightarrow
\quad
X_{\rm H}
=
\frac{\gn\,\bar M}{r(\bar R_{\rm H})}
=
1
\ ,
\label{GRH}
\ee
while the Buchdahl limit, where the central pressure becomes infinite, is located at 
\be
r(\bar R_{\rm B})
=
\frac{5}{4}\,\gn\,\bar M
\quad
\Rightarrow
\quad
X_{\rm B}
=
\frac{\gn\,\bar M}{r(\bar R_{\rm B})}
=
\frac{4}{5}
\ .
\label{GRB}
\ee
The relation~\eqref{trr} can be inverted and, at order $\gn$, yields
\be
\bar r
\simeq
r
\left[
1
+
\left(
3
-
\frac{r^2}{R^2}
\right)
\frac{\gn\,\bar M}{2\,R}
\right]
\ ,
\label{hr}
\ee
where we used $\bar R=R+\gn\,\bar M$.
We can finally replace the areal coordinate in Eq.~\eqref{Vgr2} with the expression~\eqref{hr}
and obtain the weak-field potential for $0\le \bar r< \bar R$ (corresponding to $0\le r< R$)
to second order in $\gn$ as
\be
V_{\rm GR}
&\!\!\simeq\!\!&
V_{\rm N}
+
\left(
15
-
6\,\frac{r^2}{R^2}
-\frac{r^4}{R^4}
\right)
\frac{\gn^2\,\bar M^2}{8\,R^2}
\nonumber
\\
&\!\!\simeq\!\!&
V_{\rm N}
+
\left(
15
-
6\,\xi^2
-
\xi^4
\right)
\frac{X^2}{8}
\ ,
\label{VGR}
\ee
where the Newtonian potential is now properly expressed in terms of the harmonic coordinate as 
\be
V_{\rm N}
\simeq
-\left(3-\frac{r^2}{R^2}\right)
\frac{X}{2}
\ ,
\ee
with $X=\gn\,\bar M/R$ and $\xi=r/R$.
Finally, the pressure in harmonic coordinates reads
\be
p
&\!\!\simeq\!\!&
\left(
1
-
\frac{r^2}{R^2}
\right)
\frac{3\,\gn\,\bar M^2}{8\,\pi\,R^4}
-
\left(
2
-
\frac{r^2}{R^2}
-
\frac{r^4}{R^4}
\right)
\frac{3\,\gn^2\,\bar M^3}{8\,\pi\,R^5}
\nonumber
\\
&\!\!\simeq\!\!&
\frac{3\,X^4}{8\,\gn^3\,\bar M^2}
\left[
\left(
1
-
\xi^2
\right)
-
\left(
2
-
\xi^2
-
\xi^4
\right)
X^2
\right]
\ .
\label{p_harmonic}
\ee
In particular, one finds
\be
p(0)
\simeq
\frac{3\,X^4}{8\,\gn^3\,\bar M^2}
\left(1-2\,X^2\right)
\ ,
\ee
which is positive only for $X^2<1/2$.
Clearly, this approximation breaks down for $X \gtrsim 0.7$, which is about the BNG
compactness for black hole formation.

\begin{thebibliography}{99}
%
%
%
\bibitem{Hawking:1973uf}
S.~W.~Hawking and G.~F.~R.~Ellis,
{\em The Large Scale Structure of spacetime},
Cambridge University Press, Cambridge (1973).
%
\bibitem{Bambi:2023try}
C.~Bambi,
{\em Regular Black Holes. Towards a New Paradigm of Gravitational Collapse},
Springer (2023)
[arXiv:2307.13249 [gr-qc]].
%
\bibitem{Casadio:2018qeh}
R.~Casadio, M.~Lenzi and O.~Micu,
Phys.\ Rev.\ D {\bf 98} (2018) 104016
[arXiv:1806.07639 [gr-qc]].
%
\bibitem{Casadio:2019cux}
R.~Casadio, M.~Lenzi and O.~Micu,
Eur.\ Phys.\ J.\ C {\bf 79} (2019)  894
[arXiv:1904.06752 [gr-qc]].
%
\bibitem{Casadio:2020kbc}
R.~Casadio and O.~Micu,
Phys. Rev. D \textbf{102} (2020) 104058
[arXiv:2005.09378 [gr-qc]].
%
\bibitem{Casadio:2020ueb}
R.~Casadio, M.~Lenzi and A.~Ciarfella,
Phys. Rev. D \textbf{101} (2020) 124032
[arXiv:2002.00221 [gr-qc]].
%
\bibitem{Casadio:2019pli}
R.~Casadio, O.~Micu and J.~Mureika,
Mod. Phys. Lett. A \textbf{35}, (2020) 2050172
[arXiv:1910.03243 [gr-qc]].
%
\bibitem{Casadio:2021gdf}
R.~Casadio, A.~Giusti, I.~Kuntz and G.~Neri,
Phys. Rev. D \textbf{103} (2021) 064001
[arXiv:2101.12471 [gr-qc]].
%
\bibitem{Casadio:2022gbv}
R.~Casadio, I.~Kuntz and O.~Micu,
Phys. Lett. B \textbf{834} (2022), 137455
[arXiv:2206.13588 [gr-qc]].
%
\bibitem{Casadio:2022pme}
R.~Casadio, I.~Kuntz and O.~Micu,
Eur. Phys. J. C \textbf{82} (2022) no.7, 609
[arXiv:2205.04926 [gr-qc]].
%
\bibitem{weinberg}
S.~Weinberg,
{\em Gravitation and Cosmology: Principles and Applications of the General Theory of Relativity},
Wiley (1972).
%
\bibitem{Arnowitt:1959ah}
R.~L.~Arnowitt, S.~Deser and C.~W.~Misner,
``Dynamical Structure and Definition of Energy in General Relativity,''
Phys. Rev. \textbf{116} (1959) 1322.
%
\bibitem{DAddio:2021xsu}
A.~D'Addio, R.~Casadio, A.~Giusti and M.~De Laurentis,
Phys. Rev. D \textbf{105} (2022) 104010
[arXiv:2110.08379 [gr-qc]].
%
\bibitem{tolman}
R.C.~Tolman,
``Relativity, thermodynamics, and cosmology,'' (Dover, 1987). 
%
\bibitem{Tolman:1939jz}
R.~C.~Tolman,
Phys. Rev. \textbf{55} (1939) 364.
%
\bibitem{Oppenheimer:1939ne}
J.~R.~Oppenheimer and G.~M.~Volkoff,
Phys. Rev. \textbf{55} (1939) 374.
%
\bibitem{Buchdahl:1959zz}
H.~A.~Buchdahl,
Phys.\ Rev.\  {\bf 116} (1959) 1027.
%
\bibitem{Stephani:2004ud}
H.~Stephani,
{\em Relativity: An introduction to special and general relativity},
Cambridge University Press, Cambridge (2004).
%
\bibitem{Misner:1964je}
C.~W.~Misner and D.~H.~Sharp,
Phys. Rev. \textbf{136} (1964), B571.
%
\bibitem{Hernandez:1966zia}
W.~C.~Hernandez and C.~W.~Misner,
Astrophys. J. \textbf{143} (1966) 452.
%
\bibitem{Schwarzschild:1916uq}
K.~Schwarzschild,
Sitzungsber. Preuss. Akad. Wiss. Berlin (Math. Phys. ) \textbf{1916} (1916) 189
[arXiv:physics/9905030 [physics]].
%
\bibitem{Schwarzschild:1916ae}
K.~Schwarzschild,
Sitzungsber. Preuss. Akad. Wiss. Berlin (Math. Phys. ) \textbf{1916} (1916) 424
[arXiv:physics/9912033 [physics.hist-ph]].
%
\bibitem{EventHorizonTelescope:2022wkp}
K.~Akiyama \textit{et al.} [Event Horizon Telescope],
Astrophys. J. Lett. \textbf{930} (2022) L12
[arXiv:2311.08680 [astro-ph.HE]].
%
%
\end{thebibliography}
\end{document}